# Power dissipation and electrical breakdown in black phosphorus

Michael Engel[1], Mathias Steiner[1,2,*], Shu-Jen Han[1], Phaedon Avouris[1]

[1] *IBM Research, Yorktown Heights, New York 10598, USA*

[2] *IBM Research, Rio de Janeiro, RJ 22290-240, Brazil*

[*]*msteine@us.ibm.com*

**ABSTRACT**

We report operating temperatures and heating coefficients measured in a multi-layer black phosphorus device as a function of injected electrical power. By combining micro-Raman spectroscopy and electrical transport measurements, we have observed a linear temperature increase up to 600K at a power dissipation rate of 0.896K$\mu$m$^3$/mW. By further increasing the bias voltage, we determined the threshold power and temperature for electrical breakdown and analyzed the fracture in the black phosphorus layer that caused the device failure by means of scanning electron microscopy and atomic force microscopy. The results will benefit the research and development of electronics and optoelectronics based on novel two-dimensional materials.

Black phosphorus[1,2], a layered semiconducting material with thickness-dependent electronic and optical properties is currently under intense investigation for future applications in electronics and opto-electronics.[1,2] Multi-layered black phosphorus stands out for its ambipolar electrical transport properties,[3,4] high carrier mobilities,[5,6] and a thickness-dependent bandgap[5-7] in combination with broadband photo-detection capabilities,[8-10] Its susceptibility to degradation upon exposure to water and oxygen has been studied and methods for effective surface passivation have been developed.[11-15] It is clear that the stability and integrity of black phosphorus under electrical bias, a prerequisite for its technological applicability, also needs to be verified. An examination of power dissipation and the temperatures that develop in functioning black phosphorus devices is, however, lacking.

In this Letter, we use anti-Stokes/Stokes Raman spectroscopy in order to determine operation temperatures of a multi-layer black phosphorus device under controlled variation of electrical bias. We calibrate the shifts of the principal Raman bands to the measured temperature in order to extract electrical heating coefficients. Moreover, we monitor the electrical device breakdown and analyze the resulting black phosphorus multilayer fracture within the device by means of optical microscopy, electron microscopy, and atomic force microscopy.

For device manufacturing, we utilized an exfoliated black phosphorus multi-layer on top a substrate consisting of a glass coverslip (170μm) coated with ITO (200nm) and $Al_2O_3$ (100nm) as shown in the wide-field microscope image Fig.1(a). Electronic devices were fabricated by defining contacts made of Ti (1nm), Pd (20nm), Au (40nm) layers at various positions along the flake, see Fig.1(b). The black phosphorus device was then connected with metallic probes to an electronic control and measurement system and mounted on a sample holder that positions the device with respect to the optical measurement system as sketched in Fig.1(c). An immersion-assisted, inverted optical microscope was used for micro-spectroscopy of the black phosphorus device from the underside through the optically transparent device stack. Fig.1d shows a confocal microscope image of the black phosphorus device having a channel length of 1.5micron and a channel width of 0.9micron. The image was acquired by raster scanning the device with respect to a focused laser beam ($\lambda_{laser}$=532nm) and collecting the elastically scattered laser light by means of a single photon avalanche detector. For performing Raman spectroscopy, the device was positioned such that the laser is focused onto the device center. The scattered light was filtered by a holographic notch filter and analyzed by a spectrograph equipped with a grating having 300 grooves/mm and a nitrogen cooled CCD camera. The experimental setup allows for acquisition of Raman spectra from within the device channel while the black phosphorus multilayer is electrically biased to feature a steady electrical current, which is monitored throughout the measurement.

In Fig. 2(a),(b) we show the electrical transport characteristics of the device. For comparison, both the electrical device current $I_{ds}$ and the electrical power $P_{el}=V_{ds} \cdot I_{ds}$ are plotted as a function of source drain bias $V_{ds}$. The effect of an electrostatic gate field on electrical transport is insignificant, as expected for a 50nm thick layer of black phosphorus. In Figs. 2(c),(d), we show Raman spectra that were measured at the steady-state bias levels indicated in Figs. 2(a),(b) by colored circles. The Raman spectra exhibit anti-Stokes and Stokes bands associated with the high-frequency intra-layer phonon modes $A_g^1$, $B_{2g}$, and $A_g^2$ which are spectrally located at around ±365cm$^{-1}$, ±442cm$^{-1}$, and ±470cm$^{-1}$, respectively.[16] We observe that, as function of electrical bias, the intensities and spectral positions of the three Raman bands change gradually. Since the application of external electric fields for gate voltages of up to 20V does not induce any measurable variations in the measured Raman spectrum, we conclude that the spectral effects observed here are due to electrical heating, an effect previously observed in high-frequency phonons of graphene.[17]

In order to quantify the effect of electrical current on temperature, we plot in Fig. 3(a) for each Raman transition the anti-Stokes-to-Stokes intensity ratio. The intensity ratios have been determined based on the analysis of the spectra shown in Fig.2(c),(d). We observe that for all three Raman transitions the anti-Stokes-to-Stokes intensity ratio increases as function of electrical power with similar slopes, see Fig. 3(b), evidencing an equilibrated phonon bath. By using the equation[18]

$$T_{ph} = T_0 \cdot \left[1 - \frac{k_B T_0}{E_{ph}} \ln\left(\frac{\left(I_{AS}/I_S\right)_T}{\left(I_{AS}/I_S\right)_{T_0}}\right)\right]^{-1} \qquad (1)$$

with ambient temperature $T_0$, phonon energy $E_{ph}$, and Boltzmann constant $k_B$, we transform the experimental anti-Stokes-to-Stokes intensity ratios into phonon temperatures $T_{ph}$. By analyzing the thermal radiation contribution in the measured spectra, we find that the electrical charge carriers in the black phosphorus multilayer display consistent temperatures and we conclude that phonons and charge carriers are in thermal equilibrium.

As a key result, we plot in Fig. 3(c) the operating temperatures of the black phosphorus multi-layer device as a function of injected electrical power. A linear fit in agreement with the expectation of Joule heating is obtained, and taking account for the actual dimensions of the black phosphorus multilayer (length=1500nm, width=900nm, height=50nm), we extract an electrical power dissipation rate of 0.896K$\mu m^3$/mW

In Figs. 3(c),(d), we plot the spectral peak positions extracted from the measured Raman spectra as function of electrical power. For all three Raman bands, we observe phonon softening and the slopes extracted from linear fits to the data are $\Delta(A_G^1)$=0.18±0.05cm$^{-1}$/mW, $\Delta(B_{2G})$= 0.44±0.05cm$^{-1}$/mW, and $\Delta(A_G^2)$= 0.50±0.05cm$^{-1}$/mW. Based on the observation that the three Raman bands do not display measurable spectral shifts upon application of external electric (gate) fields, we conclude that the observed shifts are solely temperature-induced. Consequently, we scale for each band the measured Raman shifts in Fig. 3(c),(d) to the experimental temperatures shown in Fig. 3(b). As a result, we obtain the following electrical heating coefficients for the three Raman-active phonon modes; $g(A_G^1)$=0.013cm$^{-1}$/K, $g(B_{2G})$= 0.033cm$^{-1}$/K, and $g(A_G^2)$= 0.038cm$^{-1}$/K. These coefficients are of the same order of magnitude as those reported for experiments conducted under thermal heating/cooling conditions.[19-21] The above values characterize a bulk-like black phosphorus multilayer, however, modifications could arise in very thin samples due to an increased band gap.

We now investigate the electrical breakdown in the black phosphorus device. At the highest bias point used in this study, $V_{ds}$=4V, $I_{ds}$=8.61mA as indicated by open symbols in Fig. 2(a),(b), we observe that the electrical device current suddenly drops. From the data point visualized by the open square in Fig.3(b), at a breakdown threshold power of $P_{BD}$=34.4mW, we derive a breakdown threshold temperature of $T_{BD}$=757K. At such elevated temperatures device breakdown in addition to structural decomposition may also involve a phase transition from black to red phosphorus[22]. In order to confirm the nature of the device failure we successively perform optical microscopy, electron microscopy, and atomic force microscopy on the device. In Fig.4(b), the scanning electron microscope image reveals the device fracture across the entire device channel with an orientation parallel to the contacts, having an approximate width of 50nm as independently revealed by the atomic force microscope measurement shown in Fig.4(b). The width of the fracture is well below the resolution limit of the optical microscope (~250nm). However, in the elastic laser scattering image of the device taken after device breakdown, see Fig.4(c), the narrow fracture reveals itself by a reduced optical scattering contrast close to one of the contacts as can be seen by comparison with the elastic scattering image of the device before breakdown, see Fig.1(d).

In summary, we have measured operating temperatures in a multi-layer black phosphorus device as a function of dissipated electrical power and we have determined electrical heating coefficients by scaling the shifts of principal Raman bands to the operation temperatures determined based on anti-Stokes-Stokes Raman thermometry. Also, we have provided experimental thresholds for electrical breakdown in multi-layer black phosphorus. Future research is needed to spatially resolve the two-dimensional heat distributions in functional black phosphorus devices to reveal the local temperature gradients and device heterogeneities that cause electric device failure.

**Acknowledgment**

We thank Dr. D. B. Farmer, J. J. Bucchignano, and B. A. Ek (all IBM Research) for expert technical assistance.

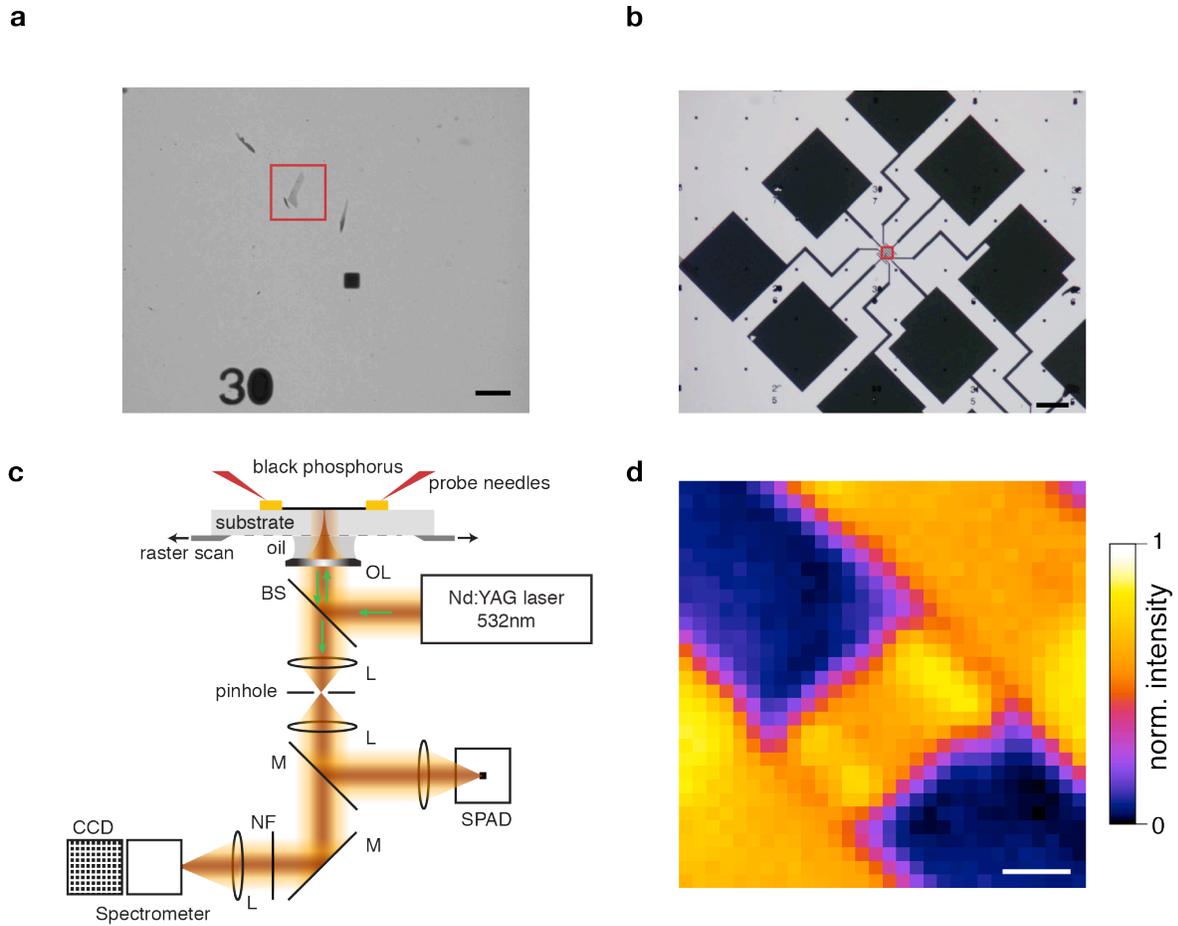

**Figure 1.** (a) Optical wide field microscope image of black phosphorus flakes exfoliated on the optically transparent substrate. The red square indicates the black phosphorus multi-layer selected for device manufaturing. Scale bar: 10 micron. (b) Optical wide field microscope image of the black phosphorus multilayer device. Scale bar: 50 micron. (c) Schematic of the optical measurement setup. OL: immersion objective, BS: beam splitter, L: lense, M: mirror, NF; notch filter, SPAD; single photon avalance detector, CCD; charge coupled array detector. (d) Confocal elastic laser scattering image of the black phosphorus multi-layer device. Scale bar: 500nm.

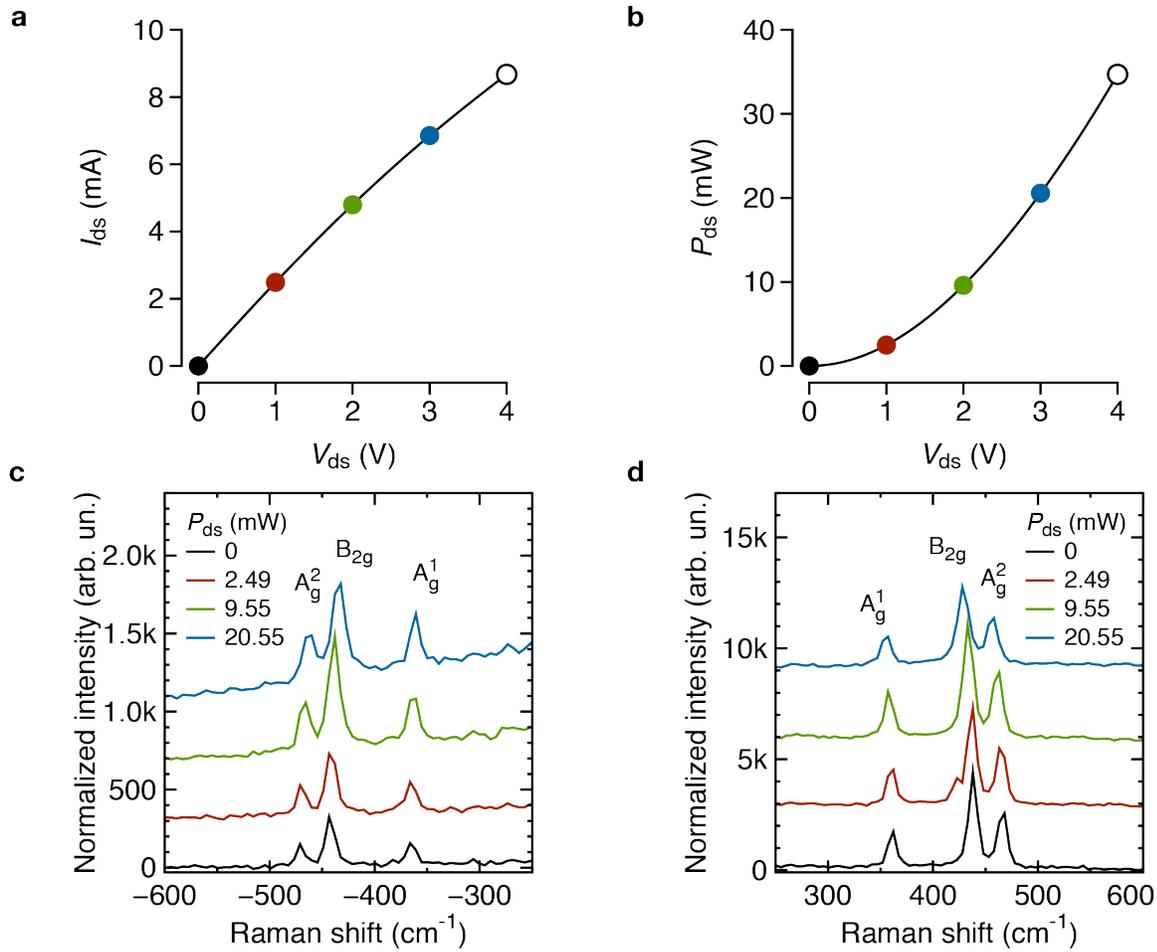

**Figure 2.** (a) Electrical output I/V-characteristic of the black phosphorus multilayer device. (b) Electrical power level in the device as function of bias voltage applied. The colored circles indicate the steady state bias points used for Raman spectra acquisition. (c) Anti-Stokes and (d) Stokes Raman spectra of the black phosphorus multi-layer taken at various electrical power levels indicated in (a) by colored circles exhibit three principal Raman bands associated with the high-frequency intralayer phonon modes $A_g^1$, $B_{2g}$, and $A_g^2$. The Raman spectra are offset vertically in order to improve visibility.

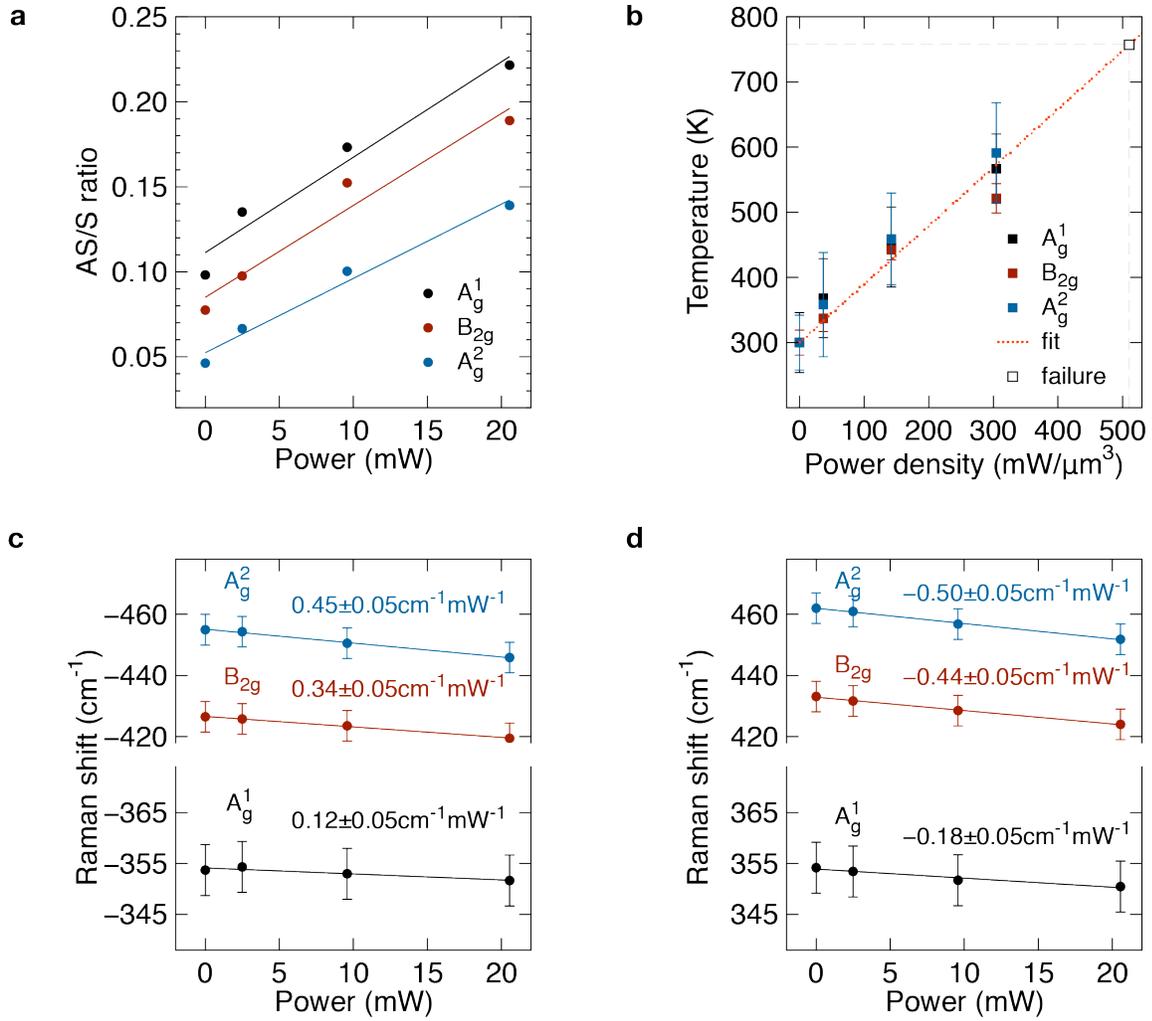

**Figure 3.** (a) Anti-Stokes-to-Stokes intensity ratio for three phonon modes determined from the black phosphorus multi-layer Raman spectra shown in Fig. 2. as function of injected electrical power. (b) Temperatures of the black phosphorus multi-layer as function of electrical power. The linear fit to the data is used to estimate the breakdown temperature of the device indicated by the open square symbol. (c), (d) Spectral shifts determined from the black phosphorus multi-layer Raman spectra shown in Fig. 2. as function of injected electrical power. The respective slopes of shifts derived from linear fits to the data are also indicated.

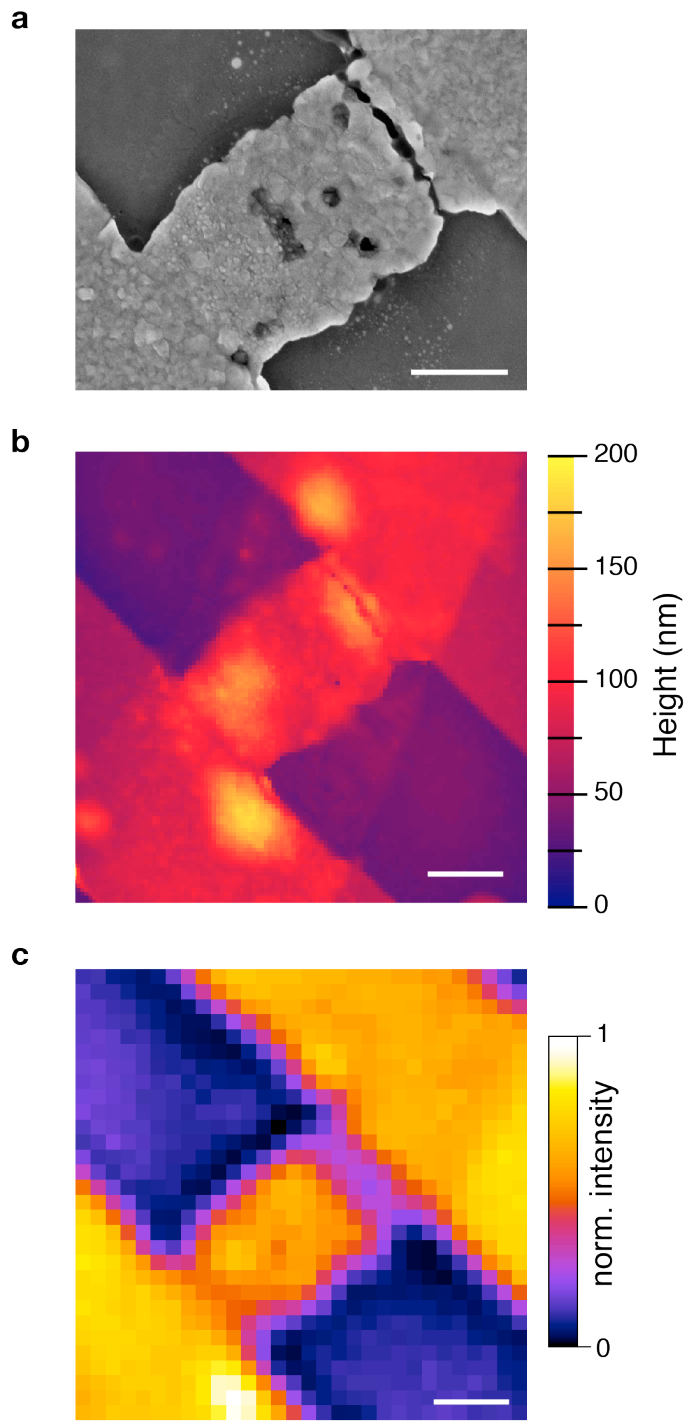

**Figure 4.** (a) Scanning electon microscopy image of the black phosphorus multi-layer device after breakdown. A fracture of the black phosphorus multi-layer can be seen close to one of the contacts. Scale bar: 500nm. (b) Atomic force microscope image of the same device. Scale bar:

1000nm. (c) Confocal elastic laser scattering microscope image of the same device. Scale bar: 1000nm.